\documentclass{aa}
\usepackage{hyperref}
\usepackage{graphicx}
\usepackage{txfonts}
\usepackage{natbib}
\bibpunct{(}{)}{;}{a}{}{,}
\usepackage{longtable}
\usepackage{lscape}
\usepackage{times}
\usepackage{textcomp}

\begin{document}

\title{Extended radio halo of the supernova remnant CTB87 (G74.9+1.2)}
%
\author{Wolfgang~Reich\inst{1}, Patricia~Reich\inst{1} and Roland Kothes \inst{2} }
\titlerunning{Halo of CTB87}
\authorrunning{W. Reich et al.}


\institute{
  Max-Planck-Institut f\"{u}r Radioastronomie, Auf dem H\"{u}gel 69,
  53121 Bonn, Germany \\{e-mail: wreich@mpifr-bonn,mpg.de; preich@mpifr-bonn.mpg.de} \and National Research Council of
  Canada, Dominion Radio Astrophysical Observatory, P.O. Box 248,
  Penticton BC, V2A 6J9, Canada \\{e-mail: Roland.Kothes@nrc-cnrc.gc.ca}}

\date{Received; accepted}

\abstract
{Breaks in the radio spectra of supernova remnants (SNRs) reflect the maximum energy of either shock-accelerated electrons 
or - in the case of pulsar wind nebulae - of electrons injected by the central pulsar. Otherwise, the break may result 
from energy losses due to synchrotron aging or it is caused by energy-dependent diffusion. A spectral steepening of the plerionic SNR CTB87 at around 11~GHz was observed in the eighties, but a recent analysis 
of CTB87's energetic properties based on new radio data raised doubt on it. CTB87 consists of a central compact component 
surrounded by a diffuse centrally peaked almost circular halo. Missing faint halo emission due to insufficient sensitivity of early high-frequency 
observations may be be the reason for the reported spectral break.} 
{We intend to clarify the high-frequency spectrum of CTB87 by new sensitive observations.  }
{We used the broad-band $\lambda2$\ cm receiver at the Effelsberg 100-m telescope for sensitive continuum observations of CTB87 and its halo 
in two frequency bands.}
{The new $\lambda2$\ cm maps of CTB87 show halo emission with a diameter of about 17$\arcmin$ or 30~pc for a distance of 6.1~kpc in agreement 
with lower-frequency data. The measured flux densities are significantly higher than those reported earlier.}
{The new $\lambda2$\ cm data establish the high-frequency continuation of CTB87's low-frequency spectrum. Any significant high-frequency spectral 
bend or break is constrained to frequencies well above about 18~GHz. The extended halo of CTB87 has a faint counterpart in $\gamma$-rays (VER J2016+37)
and thus indicates a common origin of the emitting electrons.
}


\maketitle
\section{Introduction}

CTB87 (G74.9+1.2) is an unusual plerionic supernova remnant (SNR) in the radio range. It consists of a compact kidney-shaped component called 
the `relic'-component following \citet{Kothes20}. The relic-component is surrounded 
by faint centrally peaked halo-like diffuse emission, called the 'halo' further on. CTB87 was classified as a filled-centred SNR, where 
only recently an indication of a partially polarised outer SNR-shell was reported by \citet{Kothes20}. Its distance was determined by \citet{Kothes03}
as 6.1$\pm$0.9~kpc, which means a physical size of 1.77$\pm$0.26~pc for a 1$\arcmin$ structure. CTB87 was also studied in X-rays \citep{Matheson13, Guest20}, 
who found a compact X-ray source as the likely pulsar inside of CTB87 located offset from the radio peak. The X-ray observations
did not detect any thermal emission. \citet{Guest20} argued that the spectrum 
and the morphology of CTB87 are consistent with a $\sim$20-kyr old pulsar wind nebula (PWN) expanding into a wind-blown bubble. 

At radio frequencies, a remarkable spectral steepening above 11~GHz was reported in 1986 by \citet{Morsi86}, which was based 
on Effelsberg 32-GHz observations in conjunction with lower-frequency flux densities. This spectral break was indirectly supported by 
84-GHz observations done with the IRAM 30-m telescope by \citet{Salter89}, who did not detect any emission from CTB87 at all. Observations with 
the Arcminute Microkelvin Imager (AMI) by \citet{HurleyWalker09} at frequencies between 14~GHz and 18~GHz also supported the spectral 
break, although interferometric observations are insensitive to extended emission, so that the faint extended halo emission 
of CTB87 is missed.

A recent detailed radio study of CTB87 by \citet{Kothes20} of CTB87's radio and polarisation properties casted doubt 
on the reality of the spectral break caused by synchrotron cooling, which requires an unreasonably high magnetic field energy.  One explanation 
for missing extended emission may be insufficient sensitivity or receiver stability of early 
mm observations. Additionally, in some cases, the size of the early maps
was not large enough to accurately trace faint extended emission from the outskirts of CTB87. 
Thus, before discussing alternative physical explanations for the spectral break, new high-frequency observations with improved sensitivity 
for faint large-scale emission are required to prove the early high-frequency results. 
The Effelsberg 100-m telescope is equipped with a sensitive broad-band receiver
at $\lambda2$\ cm, which seems well suited for the purpose to clarify CTB87's high-frequency spectrum.
In Sect.~2, we describe the $\lambda2$\ cm radio observations with the Effelsberg 100-m telescope, the data reduction and calibration.
Section~3 presents the resulting maps, flux integration and the derived spectra. In Sect.~4, we discuss the spectral properties of CTB87. 
Section~5 summarises our results. 

\section{Effelsberg $\lambda$2\ cm observations and data reduction}

We used the broad-band $\lambda$2\ cm (Ku-band) receiver in the secondary focus of the Effelsberg 100-m telescope during 
three almost clear nights in May 2020 and observed CTB87 with the aim to accurately trace its halo emission. The receiving system 
is mounted in the secondary focus of the telescope and consists
of two feeds with two amplifiers each for the left- and right-handed circular components. The feeds are separated by $3\farcm984$ in azimuth  
and require `software beam-switching' \citep{Morsi86} when weather effects must be removed in case they influence the astronomical signal. The 
four cooled HEMT amplifiers cover the frequency interval from 12~GHz to 18~GHz. This 6-GHz wide band is split into two 2.5-GHz wide frequency bands 
centred at 14.25~GHz and 16.75~GHz, respectively. Receiver tests\footnote{\url{www.mpifr-bonn.mpg.de/effelsberg/astronomers}} 
have shown that the sensitivity across the bandpass increases with frequency. Unfortunately, no polarimeter is connected to the receiver.
We found effective centre frequencies of 14.7~GHz and 17.3~GHz for the two frequency bands we observed. The backend provides the two 
circularly polarised signals for the two feeds and for each frequency band, which represent the total-intensity signal when circular 
polarisation is negligible, as it is the case. We measured a 
beam-width (HPBW) of 51$\arcsec$ and 43$\arcsec$ from the unresolved 
calibration sources 3C286 and 3C48 at 14.7~GHz and 17.3~GHz, respectively.
These measured HPBWs agree with those expected for the Effelsberg 100-m telescope for frequencies of 14.7~GHz and 17.3~GHz.
A summary of the observational parameters is listed in Table~\ref{ObsTab}.

\begin{table*}[thp]
\caption{Observational parameters}

\label{tab1}
\vspace{-1mm}
\centering
\begin{tabular}{lrr}
\hline
\multicolumn{1}{c}{} &\multicolumn{1}{c}{}  & \multicolumn{1}{c}{}\\
Data                           \\
\hline             
Centre Frequency (effective, GHz)                  &14.7                    &17.3 \\
Bandwidth (GHz)                           &2.5                     &2.5   \\
HPBW($\arcsec$)                            &51                      &43\\
Flux densities of 3C286, 3C48 (Jy)      &3.4/1.9                        &3.0/1.6 \\
Maps - observed/usable                  & 10/5.5                       &10/4\\
rms-noise [mJy/51" beam area]         &1.1                    &1.6  \\
\hline\hline
\end{tabular}
\label{ObsTab}
\vspace{-1mm}
\end{table*}

Ten maps were observed in the Galactic coordinate system and were centred on CTB87 ($l,b$ = 74$\fdg$95, 1$\fdg$15). 
Raster scans were observed by moving the telescope either along Galactic longitude or latitude direction, where subsequent scans were separated 
by 20$\arcsec$. The size of the maps varied between 20$\arcmin \times 20\arcmin$ and 24$\arcmin \times 24\arcmin$.

The tabulated backend-data were combined into raw maps for each of the four channels for each feed by using standard procedures for continuum observations 
with the Effelsberg 100-m telescope. No weather effects influenced the data so that there was no need to apply `software beam-switching'. 
The maps were subsequently processed using the standard NOD2 reduction 
package. Radio-frequency interference (RFI) and limited receiver stability, reflecting in baseline variations, were observed for all channels 
and varied between the different observed maps. 
Thus, a careful selection of the raw maps was needed to obtain the lowest noise level for the combined map. 
The emission from the bright quasar J2015+371 at {\it l,b} = $74\fdg87, 1\fdg22$, which is located very near to the western periphery of CTB87, 
was removed by a Gaussian fit from all single maps and its sidelobe structure was partly removed in the final map, although some faint residuals remained visible. 
The eight raw maps from each observation were individually edited for spiky interference. 
Baseline curvatures were removed by applying the `pressing method' introduced by \citet{Sofue79}. The maps from 
the second feed with a constant offset in azimuth of $3\farcm984$ were individually transformed into the Galactic 
coordinate system following the procedure described by \citet{Kothes17}. We finally used the `PLAIT' procedure \citep{Emerson88} to combine 
reliable maps observed at each of the three nights. `PLAIT' provides `basket-weaving' in the Fourier domain. 55\% and 40\% of the observed maps were found 
to have sufficient quality to be used for the combination. The maps provided by `PLAIT' from the three nights were added to 
the final 14.7-GHz and 17.3-GHz map.  The quality of the combined higher-frequency map 
was lower than that of the lower-frequency map because of less useful observations (see Table~\ref{ObsTab}).

The calibration of the data relied on mapping of 3C286 and 3C48, where 3C48 served as a secondary calibrator. The assumed flux densities are listed 
in Table~\ref{ObsTab}, which we derived from the spectral-fit parameters by \citet{Ott94} and \citet{Perley17} for the present centre frequencies.

\begin{figure}
\centering
\includegraphics[angle=-90, width=0.49\textwidth]{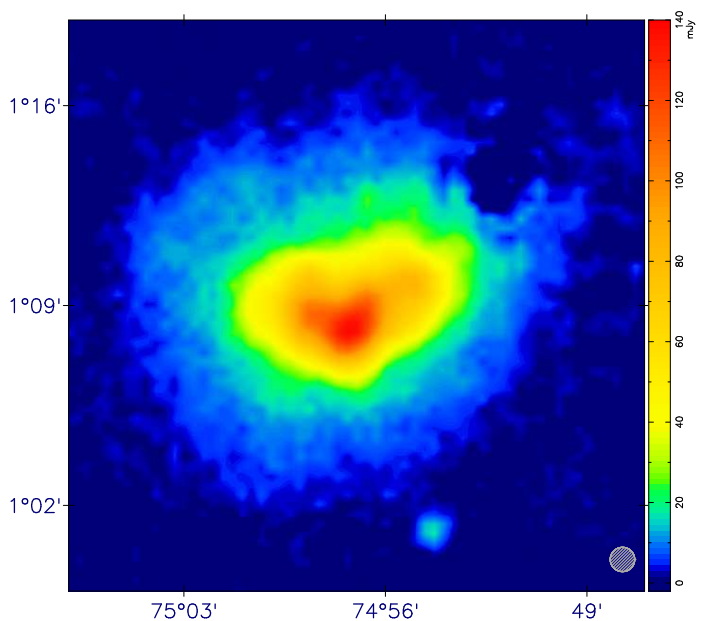}
\caption{Colour-coded Effelsberg 14.7-GHz map of CTB87 at 51$\arcsec$ angular resolution. The unrelated strong source,  
J2015+371 ({\it l,b} = $74\fdg87, 1\fdg22$), was removed from the map. The HPBW is indicated in the lower right corner.}
\label{low}
\end{figure}

\begin{figure}
\centering
\includegraphics[angle=-90, width=0.49\textwidth]{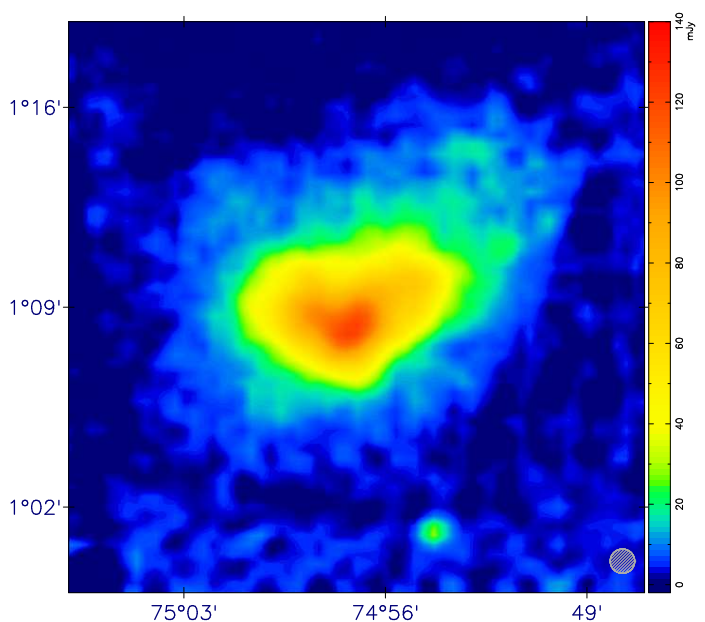}
\caption{As Fig.~1, but showing the colour-coded Effelsberg 17.3-GHz map of CTB87 smoothed to an
angular resolution of 51$\arcsec$ as of the 14.7-GHz map.}
\label{high}
\end{figure}

\section{Total-intensity maps of CTB87}

The full extent (diameter) of the halo of CTB87 is about 17$\arcmin$, so that we reduced the final map size
to 20$\arcmin \times 20\arcmin$ omitting the outer areas of the larger maps, where less observations resulted in higher noise levels. 
We noted a weak intensity gradient of the diffuse large-scale Galactic emission across the entire maps, which we removed
by subtracting a twisted plane defined by the average signal of the four $3\arcmin\times3\arcmin$-large edge-areas to improve the local zero-level 
setting of CTB87. 
The observed 14.7-GHz and 17.3-GHz total-intensity maps are shown in Figs.~\ref{low} and \ref{high} at the angular resolution of the 14.7-GHz map
of 51". We determined the rms-noise in almost flat regions of the maps and measured 1.1~mJy/beam~area and 1.6~mJy/beam~area at 14.7~GHz and 17.3~GHz, 
respectively.  

\begin{figure}
\centering
\includegraphics[angle=-90, width=0.49\textwidth]{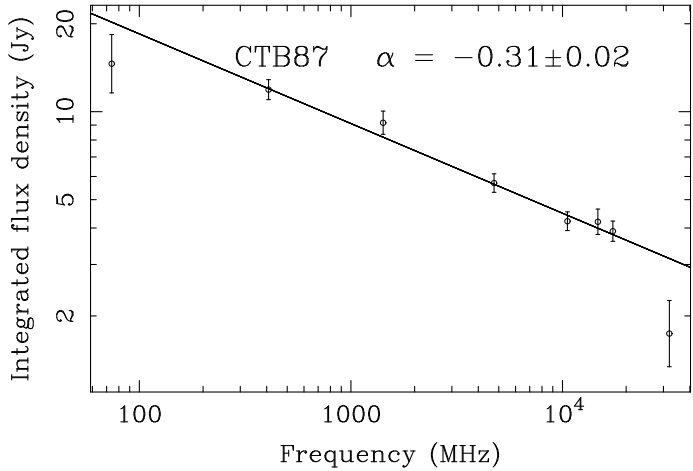}
\caption{Integrated flux-density spectrum of CTB87 including the data from
the present observations at 14.7~GHz and 17.3~GHz. Other data were taken from \citet{Kothes20} scaled to match the \citet{Perley17}
flux-density scale.}
\label{spec-all}
\end{figure}

\begin{figure}
\centering
\includegraphics[angle=-90, width=0.49\textwidth]{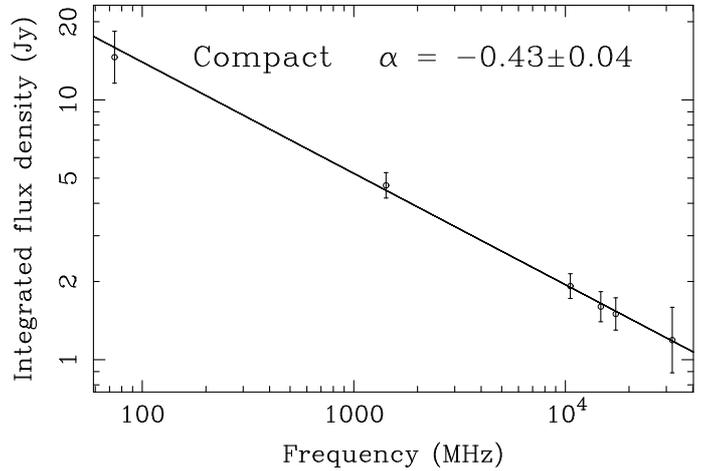}
\caption{Integrated flux-density spectrum of CTB87's compact relic-component including data from \citet{Kothes20}
on the \citet{Perley17} flux-density scale.} 
\label{spec-comp}
\end{figure}

\begin{figure}
\centering
\includegraphics[angle=-90, width=0.49\textwidth]{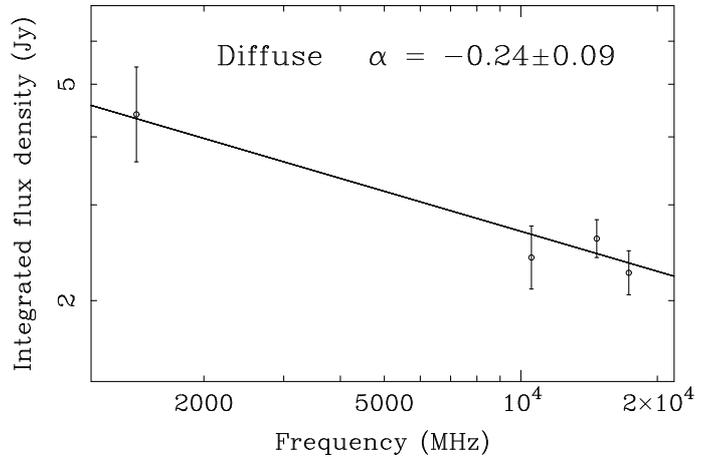}
\caption{Integrated flux-density spectrum of CTB87's halo component. The
1.4-GHz and 10.55-GHz data were taken from \citet{Kothes20} on the \citet{Perley17} flux-density scale.} 
\label{spec-bg}
\end{figure}

\subsection{Flux densities and integrated spectrum}

We integrated the total intensities  
by enclosing CTB87 and its halo by a polygon. We also performed integration in rings following \citet{Kothes20}.
Both methods gave consistent results. We measured integrated flux densities of 4.3 $\pm$ 0.4 Jy at 14.7~GHz 
and 3.9 $\pm$ 0.3 Jy at 17.3~GHz. We added these new data to the 74-MHz to 32-GHz data listed in Table~3 of \citet{Kothes20}
by replacing their too low 14.7-GHz flux density of 2.9 $\pm$ 0.4 Jy. This low flux density resulted from a too small map size 
and much less sensitivity when compared to the present map. We display the new radio spectrum of CTB87  
in Fig.~\ref{spec-all}. 
The new data indicate a continuation of the low-frequency spectrum towards frequencies of 18~GHz and
the fitted spectral index of $\alpha =-0.31 \pm 0.02$ (\it{S} \rm$\sim\nu^{+\alpha}$), agrees very well with that of \citet{Kothes20} who fitted
a spectral index of $\alpha =-0.29 \pm 0.02$ including data below ~10~GHz.

Following \citet{Kothes20}, we separated the small-scale relic-component of CTB87 from the much larger-scale 
extended halo emission via `unsharp masking' \citep{Sofue79} using the same procedure and parameters, 
i.e. smoothing the maps to $1\farcm2$ and a filter with a $2\farcm5$-wide Gaussian. We derived flux densities of
1.6 $\pm$ 0.2 Jy at 14.7~GHz and 1.5 $\pm$ 0.2 Jy at 17.3~GHz
for the relic and 2.6 $\pm$ 0.2 Jy at 14.7~GHz and 2.4 $\pm$ 0.2 Jy at 17.3~GHz for
the halo component.   
We show the overall spectrum of the relic-component of CTB87 in Fig.~\ref{spec-comp}, where a spectral index of $\alpha =-0.43 \pm 0.04$
was fitted. This, again, agrees very well with that of \citet{Kothes20} (their Fig.~6). The spectrum of the halo component
is shown in Fig.~\ref{spec-bg}, which has a spectral index of $\alpha =-0.24 \pm 0.09$. 

\subsection{Spectral indices from TT-plots}

Objects, like CTB87, with a poorly defined outer boundary may be subject to flux-density integration uncertainties 
by the zero-level setting in each map. Thus, we performed TT-plots \citep{Turtle62}, which are insensitive to zero-level settings
when comparing maps at different frequencies. It is also required that no intensity gradients are
present in different directions, which is the case for the CTB87 maps we used. The slope of the TT-plot 
gives the spectral index between the two frequencies.

We performed TT-plots using the 14.7-GHz and 17.3-GHz CTB87 maps both convolved to a common angular 
resolution of $1\farcm2$. For intensities exceeding 35(30)~mJy/beam area at 14.7~GHz(17.3~GHz), the TT-plot resulted in a spectral index 
of $\alpha =-0.49 \pm 0.02$ as shown in Fig.~\ref{tt-14-17}, which is more reliable compared to the spectral index obtained from
integrated flux densities (4.3~Jy/3.9~Jy at 14.7~GHz/17.3~GHz, see Sect. 3.1.) of $\alpha \sim -0.6$. However, the spectral 
index of $\alpha =-0.49 \pm 0.02$ is lower than the overall
spectral index of CTB87 with $\alpha =-0.31 \pm 0.02$ (Fig.~\ref{spec-all}), but close to the overall spectrum of the relic-component
(Fig.~\ref{spec-comp}). We also made TT-plots separately for the relic and the halo component at 14.7~GHz and 17.3~GHz
and obtained spectra (see Fig.~\ref{sou-2cm} and Fig.~\ref{bg-2cm}) for both components. 
The relic-component has a slightly steeper spectrum with a spectral index of $\alpha =-0.57 \pm 0.01$ (Fig.~\ref{sou-2cm}), 
while the halo component´s fit results in $\alpha =-0.15 \pm 0.02$ (Fig.~\ref{bg-2cm}) close to the overall diffuse spectral index (Fig.~\ref{spec-bg}).
These spectral indices are rather similar to those at lower frequencies and confirm the lack of any steepening
in this frequency range.

\begin{figure}
\centering
\includegraphics[angle=-90, width=0.49\textwidth]{Reich-6.ps}
\caption{Temperature-versus-temperature plot of CTB87 for the 14.7-GHz and the 17.3-GHz data
at an angular resolution of $1\farcm2$.}
\label{tt-14-17}
\end{figure}

\begin{figure}
\centering
\includegraphics[angle=-90, width=0.49\textwidth]{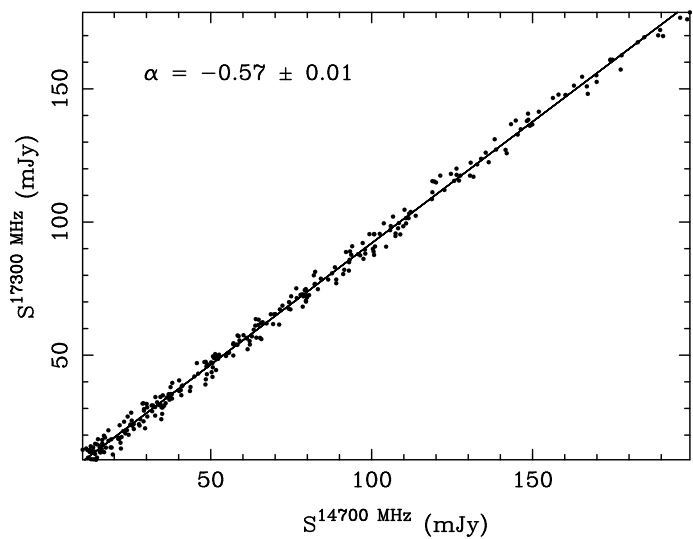}
\caption{Temperature-versus-temperature plot of CTB87 for the 14.7-GHz and 17.3-GHz 
compact component at an angular resolution of $1\farcm2$.}
\label{sou-2cm}
\end{figure}

\begin{figure}
\centering
\includegraphics[angle=-90, width=0.49\textwidth]{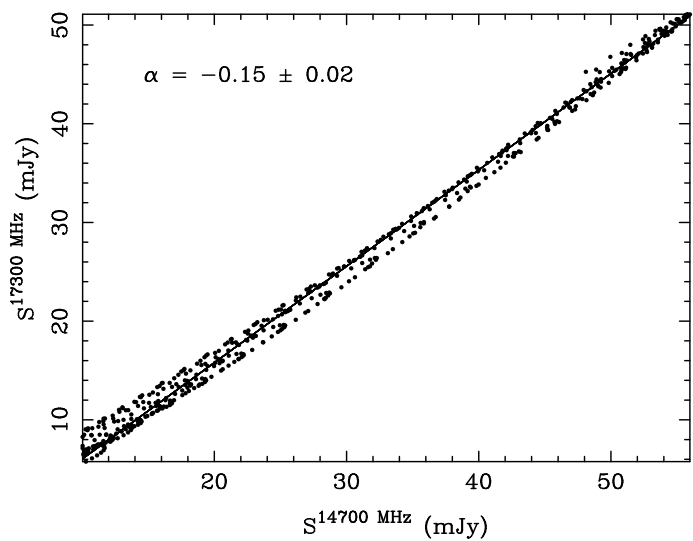}
\caption{Temperature-versus-temperature plot of CTB87 for the 14.7-GHz and 17.3-GHz 
diffuse-halo component at an angular resolution of $1\farcm2$.}
\label{bg-2cm}
\end{figure}

\begin{figure}
\centering
\includegraphics[angle=-90, width=0.49\textwidth]{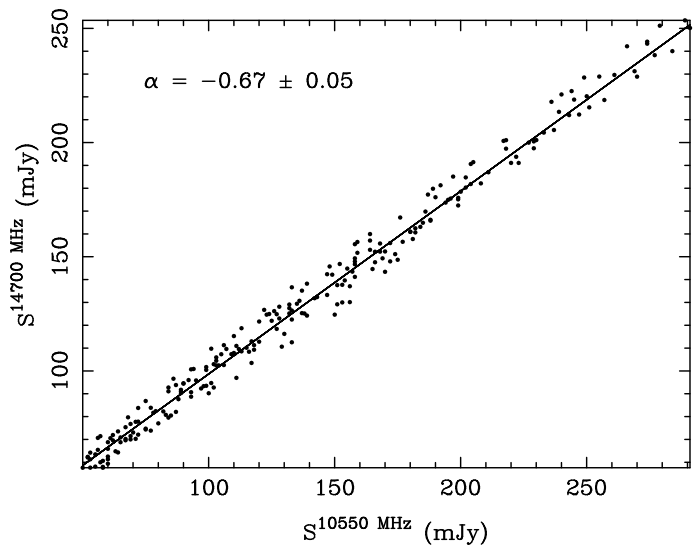}
\caption{Temperature-versus-temperature plot of CTB87 as Fig.~\ref{tt-14-17}, but now for
the 10.55-GHz and 14.7-GHz data at an angular resolution of $1\farcm2$.}
\label{tt-10-14}
\end{figure}

\begin{figure}
\centering
\includegraphics[angle=-90, width=0.49\textwidth]{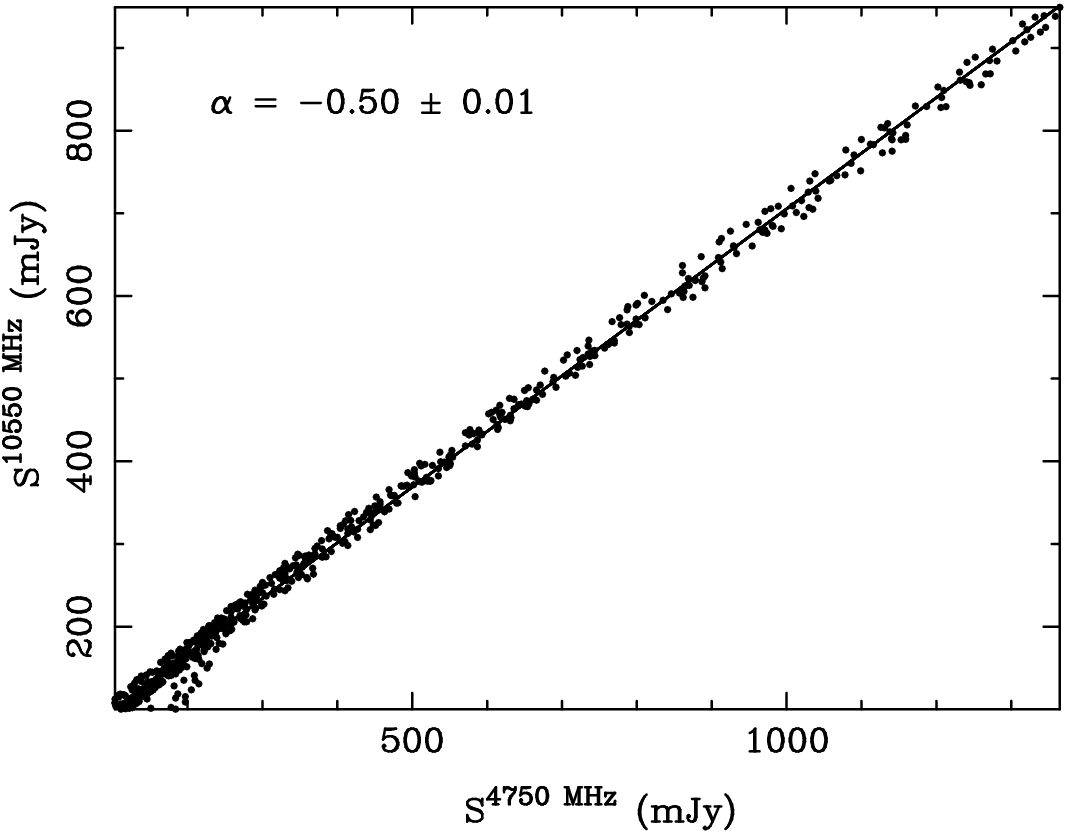}
\caption{Temperature-versus-temperature plot of CTB87 for the 4.75-GHz and 10.55-GHz
data at an angular resolution of $2\farcm5$.}
\label{tt-5-10}
\end{figure}

Effelsberg maps of CTB87 at 4.75~GHz and 10.55~GHz were presented and analysed by \citet{Kothes20}. 
They have angular resolutions of $2\farcm5$ and $1\farcm2$, respectively. We used these maps for TT-plots
in combination with the new 14.7-GHz and 17.3-GHz maps.
We present results for the combination 10.55~GHz with 14.7~GHz at $1\farcm2$ resolution in Fig.~\ref{tt-10-14} 
and between 4.75~GHz and 10.55~GHz at $2\farcm5$ resolution in Fig.~\ref{tt-5-10}. 

The spectral indices 
$\alpha$ vary slightly between the various frequency pairs from $-0.49$ to $-0.67$ and are lower
than the overall spectral index from the integrated flux densities of $\alpha = -0.31\pm0.02$ (Fig.~\ref{spec-all}). 
This indicates a dominance of the compact component of CTB87, which has a spectral index of $\alpha = -0.43\pm0.04$ 
as shown in Fig.~\ref{spec-comp} and $\alpha = -0.57\pm0.01$ from the TT-plot between 14.7~GHz and 17.3~GHz (Fig.~\ref{sou-2cm}).
Despite these measured variations, the compact component of CTB87 has a steeper spectrum compared to the diffuse 
halo component, confirming the analysis of \citet{Kothes20} at lower frequencies.

Unfortunately, the quality of the 32-GHz map of CTB87 shown by \citet{Kothes20} is lower than that of all lower-frequency maps. 
From Fig.~\ref{spec-all} and Fig.~\ref{spec-comp}, we see that the 32-GHz flux densities are clearly below the extrapolated integrated 
flux-density spectrum, but in accordance with 
the compact component spectrum. This indicates missing diffuse-halo emission at 32~GHz, due to low
sensitivity and limited map size.
We obtained a spectral index for the compact component from the TT-plots between the 14.7-GHz 
and the 32-GHz data at an angular resolution of the 14.7-GHz map of 51$\arcsec$, which results 
in $\alpha = -0.69\pm0.23$ in rough agreement with the lower-frequency results.
The accuracy of the separation of the diffuse and compact components is limited and suffered 
from the low quality of the 32-GHz map, so that we don't discuss the results including the 32-GHz components further.  
More sensitive single-dish observations in the 30-GHz frequency range are required to study the two components in this frequency range.

As mentioned earlier, the AMI observations of CTB87 as reported by \citet{HurleyWalker09} suffer from missing large-scale emission, so that 
the listed flux densities are clearly below those measured with single-dish telescopes. From the published combined 15.75-GHz 
AMI survey map \citep{AMI13}, we measured a halo size of 12$\arcmin$, which is much smaller compared to 17$\arcmin$ measured from our 
single-dish maps. This proves that our assumption of missing flux density is correct.

\section{Discussion}

The aim of the new $\lambda$2\ cm observations was to clarify whether missing large-scale emission in early high-frequency
maps is the reason for the reported spectral break of CTB87 at around 11~GHz. This is clearly the case. The integrated high-frequency flux densities
are higher compared to previous results. The spectrum of CTB87 above 14.7~GHz continues with $\alpha \sim -0.30$ at least
to about 18~GHz.

The TT-plot spectrum of the relic-component is clearly steeper than that of the diffuse 17$\arcmin$ halo for frequencies below $\sim$ 10~GHz.
We derived a spectral index $\alpha = -0.43 \pm 0.03$ for the relic-component compared to $\alpha = -0.24 \pm 0.09$ for the diffuse outer halo, which needs explanation.

A discussion of the nature and origin of the two components of CTB87 was already presented by \cite{Kothes20}, which remains valid in view of the new observations, 
that we refer to \citet{Kothes20} for details. In brief: The steep radio spectrum of the relic-component of CTB87 has similarities to a few other PWN like DA 495 
\citep{Kothes08}, G76.9+1.0 \citep{Landecker97} or G141.2+5.0 \citep{Kothes14} and is believed to be the result of the interaction of the SNR's reverse shock with 
the PWN in the presence of a strong magnetic field. Indeed, \citet{Kothes20} estimated a strong magnetic field component along the line-of-sight in the range between 
350~$\mu$G and 580~$\mu$G for the compact component of CTB87 from an analysis of polarisation data. The shift of the 
cooling break-frequency to more than 18~GHz, however, reduces the magnetic field strength of the PWN by just ~13\%, which still implies an unreasonably high magnetic 
energy (see Fig.~11 of \citet{Kothes20}), so that any break by synchrotron cooling is expected to occur at a much higher frequency.

A flat halo spectrum is difficult to explain by shock-accelerated cosmic-ray electrons, which have a steeper spectrum, instead it resembles a typical PWN spectrum. 
The relic's spectrum is close to that expected for a shock-accelerated cosmic-ray spectrum. The relic-component is kidney-shaped with small intensity variations. 
This morphology does not indicate a SNR shock moving outwards or in case of a reverse shock moving inwards.
We note that CTB87 shows a low fraction of polarised emission at 4.75~GHz and 10.55~GHz (see \citet{Kothes20} for details) with an average fraction of the order 
of 8$\%$ and maximum values around 20$\%$ at 10.55~GHz. This indicates a dominating turbulent magnetic field with a smaller regular magnetic field component. 
The steeper spectrum of the relic-component may result from turbulent magnetic fields on small scales causing 
scattering preferably for the lower-energy electrons, so that a smaller fraction reaches the halo region when compared to the higher-energy 
electrons. This will flatten the halo spectrum as it is observed. Of course, shock-accelerated emission from a reverse shock will also enhance
the emission from the compact component and will mix with the emission originating from the pulsar-injected electrons. 
The synchrotron emissivity is clearly larger in the relic-component compared to the halo region, when we assume a symmetric source shape.
The volume ratio of the relic-component when compared with the halo component is about 1/20, while both flux densities are comparable. This means an average emissivity 
ratio of about 1/20. For the case of the same electron density in the relic and halo component of CTB87, the magnetic field in the halo drops to a 
level of about 20\% to 25\% of the relic-component´s value when taking the spectral differences of both components into account. 

\begin{figure}
\centering
\includegraphics[angle=-90, width=0.49\textwidth]{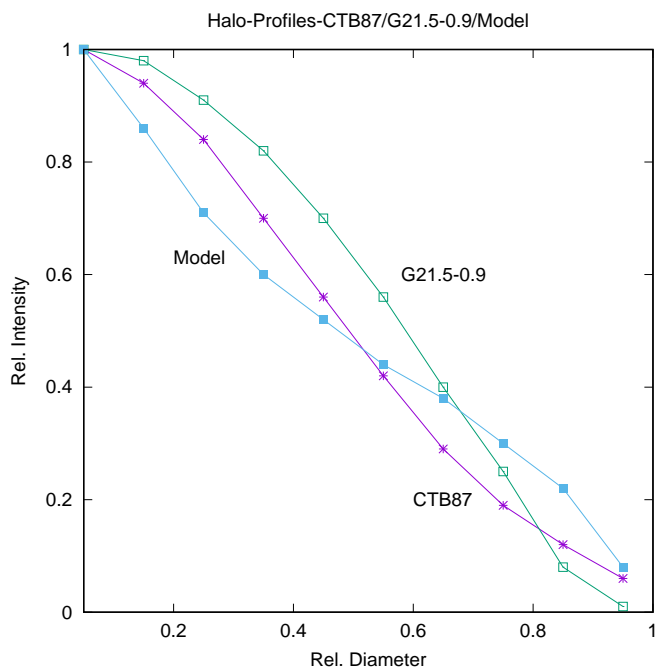}
\caption{Halo profiles of CTB87 and G21.5-0.9 compared with a halo model with constant electron density and radially decreasing magnetic field with 
$\it B_{\rm radial}$/\it B$_{\rm random}$ = \rm 0.7.}
\label{profiles}
\end{figure}

We note some similarity of the averaged radial halo-intensity of CTB87 with that of the PWN G21.5-0.9 \citep{Fuerst88} (Fig.~\ref{profiles}), although the halo of 
G21.5-09 is about 2.3~pc in size and thus much smaller than the CTB87 halo with about 30~pc. G21.5-0.9 has a compact 
central component, although of different morphology compared to CTB87, and a diffuse centrally peaked 
radio halo with a size of about 2.3~pc for a distance of 4.7$\pm$0.4~kpc \citep{Camilo06}. 
The halo of G21.5-0.9 has a radial magnetic field and a ratio $\it B_{\rm radial}/\it B_{\rm random}$ of about 0.7 to 1.
\citet{Fuerst88} could not well reconstruct the distribution of the electron density and the magnetic field 
strength of the halo of G21.5-0.9 by simple models (their Fig.~6) and concluded that G21.5-0.9 defies the explanation as an expanding synchrotron nebula. A halo 
model with a constant electron density and a linear decrease of the magnetic field strength from the centre outwards is shown in Fig.~\ref{profiles} for comparison.
This is the model with an emissivity profile closest to the observations of G21.5-0.9, although it clearly does not fit the observations 
well. This is also the case for the halo profile of CTB87, although the electron and magnetic 
field distributions seems to be even more complex when compared to G21.5-0.9.

The halos of the two PWNs G21.5-0.9 and CTB87 have morphological similarities to the TeV halos surrounding Geminga, PSR B0656+14 (located 
in the Monogem Ring) \citep{Abeysekara17} and PSR J0622+3749 \citep{Aharonian21}. 
However, the size of these TeV halos is on scales of several degrees or about 60~pc and thus about two times larger compared to the halo of 
CTB87. The TeV halos are believed to be much older than CTB87 and G21.5-0.9. \citet{Saha16} studied the $\gamma$-ray emission of VER J2016+37 being coincident with 
CTB87 and proposed a broken power-law distribution of electrons to explain the observed radio, X-ray and TeV data, while a hadronic model requires ambient 
matter densities exceeding $\sim$20 cm$^{-3}$, which is not indicated by observations and thus considered to be unlikely. 
A global study of the Cygnus region based on $\gamma$-ray data by \citet{Abeysekara18}, using VERITAS and FERMI observations, also discusses the coincident
$\gamma$-ray emission with CTB87. Interestingly, the extended diffuse $\gamma$-ray emission observed by VERITAS (\citet{Abeysekara18} their Fig.~20) has a size very 
similar to that of CTB87's radio halo suggesting a common origin of the emitting electrons. The distribution of the $\gamma$-ray emission from CTB87 suffers from 
the overlap with the 
$\gamma$-ray emission from the quasar J2015+371 next to CTB87 and the separation accuracy is limited by the low angular resolution of $\gamma$-ray telescopes,  
so that a detailed comparison is currently not meaningful. \citet{Abeysekara18} were able to fit a single power-law electron spectrum to the CTB87 $\gamma$-ray data
from VERITAS and FERMI, which contrasts to the result of \citet{Saha16}.

c

\section{Summary}

New $\lambda$2~cm observations prove that the radio spectrum of the PWN CTB87 is straight up to $\sim$ 18~GHz. These observations demonstrate 
a considerable improvement in receiver technology compared 
to early single-dish observations of CTB87 at frequencies higher than $\sim$ 10~GHz. These early maps 
and also interferometric maps suffer from insufficient sensitivity for faint extended emission. 
The flat-spectrum halo of CTB87 with a typical PWN spectrum indicates diffusion of electrons 
outwards from the central object, which has a steeper spectrum although similar to a few other PWNs. 
We note that the radial emissivity profile of the halo of CTB87 is similar to that of the PWN G21.5-0.9. 
Both profiles can not be explained by simple power-law distributions of electrons and magnetic field components.
The extent of CTB87's radio halo agrees in size with VERITAS emission at TeV energies, suggesting a common origin
from the still undetected pulsar in the centre of CTB87. 

\begin{acknowledgements}
Based on observations with the Effelsberg 100-m telescope of the MPIfR (Max-Planck-Institut f\"ur Radioastronomie).
We like to thank Rohit Dokara for careful reading of the manuscript and the referee, Miroslav D. Filipovic, for 
valuable suggestions for improvements.

\end{acknowledgements}

\bibliographystyle{aa}
\bibliography{bbfile}

\end{document}